\documentclass{llncs}
\usepackage{graphicx}
\usepackage{geometry}
\ifx\pdfoutput\@undefined\usepackage[usenames,dvips]{color}
\else\usepackage[usenames,dvipsnames]{color}
\IfFileExists{pdfcolmk.sty}{\usepackage{pdfcolmk}}{} 
\fi
\usepackage[plainpages=false,pdfpagelabels,pagebackref=false,naturalnames=true,hyperindex=true,]{hyperref}
\hypersetup{colorlinks=true,
urlcolor=Cerulean,linkcolor=BrickRed,citecolor=RoyalBlue,
  pdfpagemode=None,
  pdfstartview=FitH}
\usepackage[all]{hypcap}

\usepackage{booktabs} 
\usepackage[table]{xcolor}
\usepackage{longtable}

\begin{document}
%
\title{Measuring Complexity in an Aquatic Ecosystem}

\author{Nelson Fern\'{a}ndez\inst{1},\inst{2}\and Carlos Gershenson\inst{3},\inst{4}}
\institute{
Laboratorio de Hidroinform\'{a}tica, Facultad de Ciencias B\'{a}sicas\\
Univesidad de Pamplona, Colombia\\
\email{nfernandez@unipamplona.edu.co} \ 
\url{http://unipamplona.academia.edu/NelsonFernandez}\\
\and
 Centro de Micro-electr\'onica y Sistemas Distribuidos,\\
Universidad de los Andes, M\'erida, Venezuela\\
\and
 Departamento de Ciencias de la Computaci\'on\\
Instituto de Investigaciones en Matem\'aticas Aplicadas y en Sistemas \\
Universidad Nacional Aut\'onoma de M\'exico\\
\email{cgg@unam.mx} \
\url{http://turing.iimas.unam.mx/~cgg}\\
\and Centro de Ciencias de la Complejidad \\
Universidad Nacional Aut\'onoma de M\'exico
}

\maketitle              

\begin{abstract}
We apply formal measures of emergence, self-organization, homeostasis,
autopoiesis and complexity to an
aquatic ecosystem; in particular to the physiochemical component of an
Arctic lake. These measures are based on information theory. Variables with an homogeneous
distribution have higher values of emergence, while variables with a more heterogeneous distribution have a higher self-organization. Variables with a high complexity reflect a balance between change (emergence) and regularity/order (self-organization). In addition, homeostasis values coincide with the variation of the winter and summer seasons. Autopoiesis values show a higher
degree of independence of biological components over their environment. Our approach shows how the ecological dynamics can be described in terms of information.

\keywords{Complex Systems, Information Theory, Complexity, Self-organization, Emergence, Homeostasis, Autopoiesis}
\end{abstract}

\section{Introduction} 

Water bodies have always been relevant. In particular, lakes provide  a broad source of water, food, and recreation. Arctic lakes are one of the most vulnerable aquatic ecosystems on the planet since they are changing rapidly,  due to the effects of global warming.

 The water column (limnetic zone) of an Arctic lake is well-mixed; this means that there are no layers with different temperatures. During winter, the surface of the lake is ice covered. During summer, ice melts and the water flow and evaporation increase. Consequently, the two climatic periods (winter and summer) in the Arctic region cause a typical hydrologic behavior in lakes. This behavior influences the physiochemical subsystem of the lake. One or more components or subsystems can be an assessment for the Arctic lakes dynamics, for example: physiochemical, limiting nutrients and photosynthetic biomass for the planktonic and benthic zones. 

In recent years, the scientific study of complexity in ecological systems, including lakes, has increased the understanding of a broad range of phenomena, such as diversity, abundance, and hierarchical structure~\cite{lizcano2004using}. It is important to consider that lakes exhibit properties like emergence, self-organization, and life. Lake dynamics generate novel information from the relevant interactions among components. Interactions determine the future of systems and their complex behavior. Novel information limits predictability, as it is not included in initial or boundary conditions. It can be said that this novel information is emergent since it is not in the components, but produced by their interactions. Interactions can also be used by components to self-organize, i.e.\ produce a global pattern from local dynamics. The balance between change (chaos) and stability (order) states has been proposed as a characteristic of complexity~\cite{Langton1990,Kauffman1993}. Since more chaotic systems produce more information (emergence) and more stable systems are more organized, complexity can be defined as the balance between emergence and self-organization. In addition, there are two properties that support the above processes: homeostasis refers to regularity of states in the system and autopoiesis that reflects autonomy.

Recently, abstract measures of emergence, self-organization, complexity, homeostasis and autopoiesis based on information theory have been proposed~\cite{GershensonFernandez:2012,Fernandez2013Information-Mea}, with the purpose of clarifying their meaning with formal definitions. In this work, we apply these measures to an aquatic ecosystem. The aim of this application to an Arctic lake is to clarify the ecological meaning of these notions, and to show how the ecological dynamics can be described in terms of information. With this approach, the complexity in biological and ecological systems can be studied.

In the next section, we present a brief explanation of measures of self-organization, emergence, complexity, homeostasis, autopoiesis. Section~\ref{sec:results} describes our experiments and results with the Arctic lake, which illustrate the usefulness of the proposed measures, closing with conclusions in Section\ref{sec:conclusions}.

\section{Measures}

Emergence refers to properties of a phenomenon that are present in one description and were not in another description.  In other words, there is emergence in a phenomenon information is produced.
Shannon~\cite{Shannon1948} proposed a quantity which measures how much information was produced by a process. Therefore, we can say that the emergence is the same as the Shannon's information $I= -K \sum_{i=i}^{n} p_{i} \log p_{i}$ where K is a positive constant and $p_{i}$ is the probability of a symbol from a finite alphabet from appearing in a string. Thus $E=I$.

Self-organization has been correlated with an increase in order, i.e. a reduction of entropy~\cite{GershensonHeylighen2003a}. If emergence implies an increase of information, which is analogous to entropy and disorder, self-organization should be anti-correlated with emergence. We propose as the measure $S = 1 - I = 1 - E$.

We can define complexity $C$ as the balance between change (chaos) and stability (order). We can use emergence and self-organization which respectvely measure that.  Hence we propose: $C = 4 \cdot E \cdot S$.  The constant 4 is added to normalize the measure to $[0, 1]$.

For homeostasis $H$, we are interested on how all variables of a system change or not in time. A useful function for comparing strings of equal length is the Hamming distance. The normalized Hamming distance $d$ measures the percentage of different symbols in two strings $X$ and $X'$. Thus, $1-d$ indicates how similar two strings are. To measure $H$, we take the average of these state similarities.

As it has been proposed, adaptive systems require a high $C$ in order to be able to cope with changes of its environment while at the same time maintaining their integrity~\cite{Langton1990,Kauffman1993}.    $X$ can represent the trajectories of the variables of a system and $Y$ can represent the trajectories of the variables of the environment of the system.  If $X$ has a high $E$, then it would not be able to produce its own information. With a high $S$, $X$ would not be able to adapt to changes in $Y$. We propose $A = \frac{C(X)}{C(Y)}$, so that higher values of $A$ indicate a higher $C$ of a system relative to their environment. 

Details of these measures can be found in~\cite{Fernandez2013Information-Mea}.

\section{Results}
\label{sec:results}

The data from an Arctic lake model used in this section was obtained using The Aquatic Ecosystem Simulator~\cite{Randerson2008}. Table 1  shows the variables and daily data we obtained from the Arctic lake simulation. The model used is deterministic, so there is no variation in different simulation runs. There are a higher dispersion for variables such as temperature ($T$) and light ($L$) at the three zones of the Arctic lake (surface=$S$, planktonic=$P$ and benthic=$B$; Inflow and outflow ($I\&O$), retention time ($ RT$) and evaporation ($Ev$) also have a high dispersion, $Ev$ being the variable with the highest dispersion.

\begin{table}[htbp]
  \centering
  \caption{Physiochemical variables considered in the Arctic lake model.}
    {\footnotesize \begin{tabular}{|l|l|l|l|l|l|l|l|}
    \toprule
\toprule
    \textbf{Variable} & \textbf{Units} & \textbf{Acronym} & \textbf{Max} & \textbf{Min} & \textbf{Median} & \textbf{Mean}  & \textbf{std.\ dev.} \\
    \midrule
    Surface Light & MJ/m2/day & $SL$    & 30    & 1     & 5.1   & 11.06 & 11.27 \\
    Planktonic Ligth & MJ/m2/day & $PL$    & 28.2  & 1     & 4.9   & 10.46 & 10.57 \\
    Benthic Light & MJ/m2/day & $BL$    & 24.9  & 0.9   & 4.7   & 9.34  & 9.33 \\
    Surface Temperature & Deg C & $ST$    & 8.6   & 0     & 1.5   & 3.04  & 3.34 \\
    Planktonic Temperature & Deg C & $PT$    & 8.1   & 0.5   & 1.4   & 3.1   & 2.94 \\
    Benthic Temperature & Deg C & $BT$    & 7.6   & 1.6   & 2     & 3.5   & 2.29 \\
    Inflow and Outflow & m3/sec & $I\&O$  & 13.9  & 5.8   & 5.8   & 8.44  & 3.34 \\
    Retention Time & days  & $RT$    & 100   & 41.7  & 99.8  & 78.75 & 25.7 \\
    Evaporation & m3/day & $Ev$    & 14325 & 0     & 2436.4 & 5065.94 & 5573.99 \\
    Zone Mixing & \%/day & $ZM$    & 55    & 45    & 50    & 50    & 3.54 \\
    Inflow Conductivity & uS/cm & $ICd$   & 427   & 370.8 & 391.4 & 396.96 & 17.29 \\
    Planktonic Conductivity & uS/cm & $PCd$   & 650.1 & 547.6 & 567.1 & 585.25 & 38.55 \\
    Benthic Conductivity & uS/cm & $BCd$   & 668.4 & 560.7 & 580.4 & 600.32 & 40.84 \\
    Surface Oxygen & mg/litre & $SO2$   & 14.5  & 11.7  & 13.9  & 13.46 & 1.12 \\
    Planktonic Oxygen & mg/litre & $PO2$   & 13.1  & 10.5  & 12.6  & 12.15 & 1.02 \\
    Benthic Oxygen & mg/litre & $BO2$   & 13    & 9.4   & 12.5  & 11.62 & 1.51 \\
    Sediment Oxygen & mg/litre & $SdO2$ & 12.9  & 8.3   & 12.4  & 11.1  & 2.02 \\
    Inflow pH & ph Units & $IpH$   & 6.4   & 6     & 6.2   & 6.2   & 0.15 \\
    Planktonic pH & ph Units & $PpH$   & 6.7   & 6..40 & 6.6   & 6.57  & 0.09 \\
    Benthic pH & ph Units & $BpH$   & 6.6   & 6.4   & 6.5   & 6.52  & 0.07 \\
    \bottomrule
    \end{tabular}%
    }
  \label{tab:vars}%
\end{table}

\subsection{Emergence, Self-organization, and Complexity}

\begin{figure}[htbp]
\begin{center}
  \includegraphics[width=0.98\textwidth]{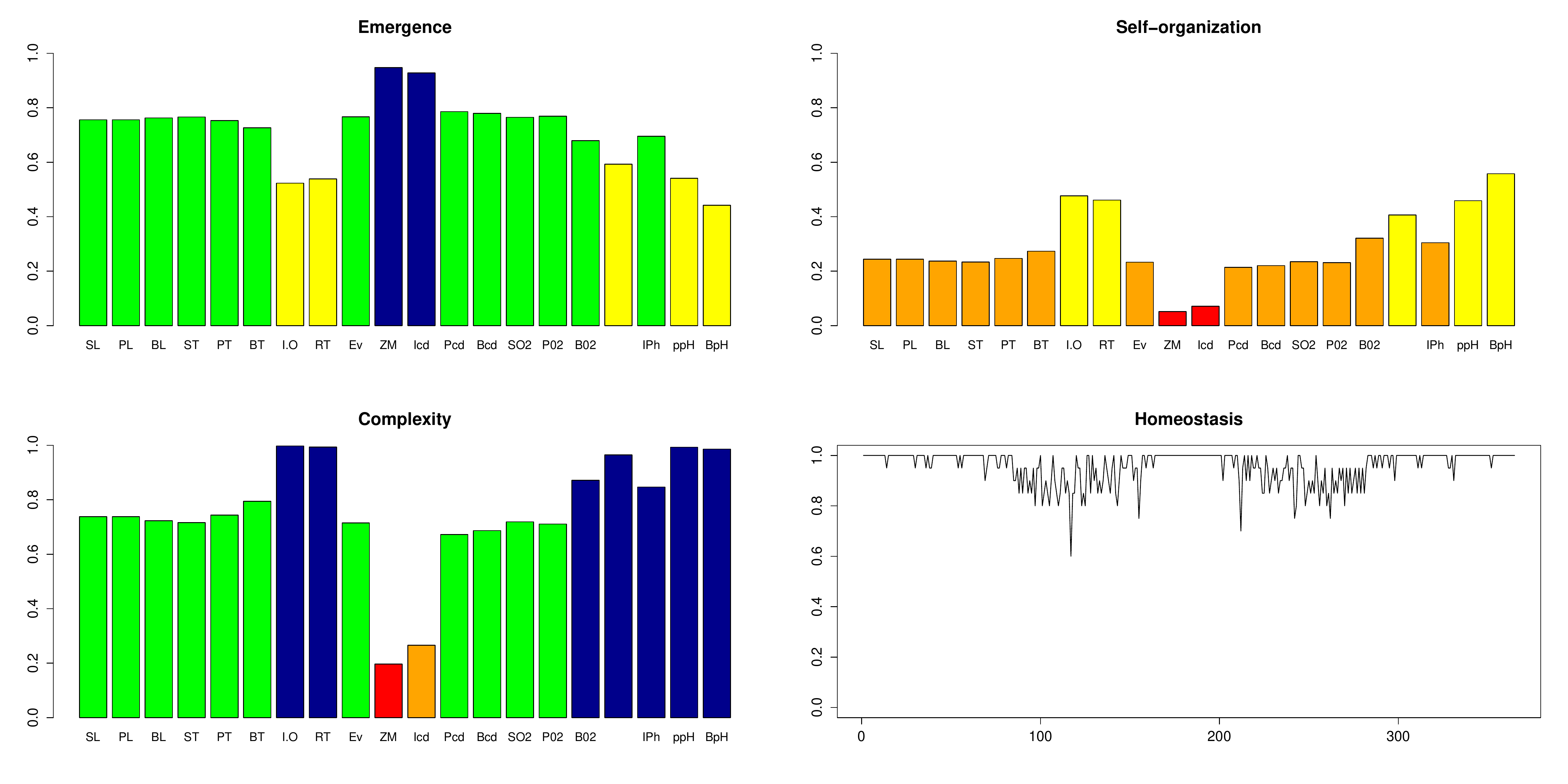}\\
\caption{$E$, $S$, and $C$ of physiochemical variables of the Arctic lake model and daily variations of homeostasis $H$ during a simulated year.}
\label{fig:ESCH}
\end{center}
\end{figure}

Figure~\ref{fig:ESCH} shows the values of $E$, $S$, and $C$ of the physiochemical subsystem\footnote{The variables were normalized to base 10 using the method described in~\cite{Fernandez2013Information-Mea}}. Variables with a high complexity $C \in  [0.8, 1]$ reflect a balance between change/chaos ($E$) and regularity/order ($S$). This is the case of benthic and planktonic $pH$ ($BpH$; $PpH$), $I\&O$ (Inflow and Outflow) and $RT$ (Retention Time). For variables with high emergencies ($E > 0.92$), like Inflow Conductivity ($ICd$) and Zone Mixing ($ZM$), their change in time is constant; a necessary condition for exhibiting chaos.  For the rest of the variables, self-organization values are low ($S < 0.32$), reflecting low regularity. It is interesting to notice that in this system there are no variables with a high $S$ nor low $E$.

Since $E, S, C \in [0, 1]$, these measures can be categorized into five categories described on the basis of an adjective, a range value, and a color for a scale from very high to very low.  The categories are: \emph{Very Low} $\in [0, 0,2]$: red, \emph{Low} $\in (0.2, 0.4]$: orange, \emph{Fair} $\in (0.4, 0,6]$: yellow, \emph{High} $\in(0.6, 0.8]$: green and \emph{Very High} $\in(0.8, 1]$: blue. This categorization is inspired on the Colombian water pollution indices. These indices were proposed by~\cite{Ramirez2003}.

We can divide the variables in the following complexity categories:

{\bf Very High Complexity} $C \in [0.8, 1]$. The following variables balance $S$ and $E$: benthic and planktonic $pH$ ($BpH$, $PpH$), inflow and outflow ($I\&O$), and retention time ($RT$). It is remarkable that the increasing of the hydrological regime during summer is related in an inverse way with the dissolved oxygen ($SO_2$; $BO_2$). It means that an increased flow causes oxygen depletion. Benthic Oxygen ($BO_2$) and Inflow $Ph$ ($IpH$) show the lowest levels of the category. Between both, there is a negative correlation: a doubling of $IpH$ is associated with a decline of $BO_2$ in 40 percent.

{\bf High Complexity} $C \in [0.6, 0.8)$. This group includes 11 of the 21 variables and involves a high $E$ and a low $S$. These 11 variables that showed more chaotic than ordered states are highly influenced by the solar radiation that defines the winter and summer seasons, as well as the  hydrological cycle. These variables were: Oxygen ($PO_2$, $SO_2$); surface, planktonic and benthic temperature ($ST$, $PT$, $BT$); conductivity ($ICd$, $PCd$, $BCd$); planktonic and benthic light ($PL$, $BL$); and evaporation ($Ev$).

{\bf Very Low Complexity} $C \in [0, 0.2)$. In this group, $E$ is high, and $S$ is very low. This category includes the inflow conductivity ($ICd$) and water mixing variance ($ZM$). Both are correlated.

\subsection{Homeostasis}

The homeostasis was calculated by comparing the variation of all variables, representing the state of the Arctic subsystem every day. The timescale is very important, because $H$ can vary considerably if we compare states every minute or every month. 
The $h$ values have a mean ($H$) of 0.957 and a standard deviation of 0.065. The minimum h is 0.60 and the maximum $h$ is 1.0. In an annual cycle, homeostasis shows four different patterns, as shown in Figure \ref{fig:ESCH}, which correspond with the seasonal variations between winter and summer. These four periods show scattered values of homeostasis as the result of transitions between winter and summer and winter back again. 

\subsection{Autopoiesis}

Autopoiesis was measured for three components (subsystems) at the planktonic and benthic zones of the Arctic lake. These were physiochemical ($PC$), limiting nutrients ($LN$) and biomass ($BM$). They include the variables and organisms related in Table \ref{tab:comps}, where the averaged $C$ of the variables is shown.

\begin{longtable}{|p{.17\linewidth}|p{.3\linewidth}|p{.06\linewidth}|p{.3\linewidth}|p{.06\linewidth}|}
  \caption{Variables and organisms used for calculating autopoiesis.}  \label{tab:comps}\\\hline
    \textbf{Component} & \textbf{Planktonic zone} & $C$ & \textbf{Benthic zone} & $C$ \\\hline
    Physiochemical & Light, Temperature, Conductivity, Oxygen, pH & 
    \cellcolor{green!50}0.771&
    Light, Temperature, Conductivity, Oxygen, Sediment Oxygen, pH &
    \cellcolor{green!50}0.861\\\hline
    Limiting Nutrients & Silicates, Nitrates, Phosphates, Carbon Dioxide &
    \cellcolor{orange!50}0.382& 
    Silicates, Nitrates, Phosphates, Carbon Dioxide &
    \cellcolor{orange!50}0.338\\\hline
    Biomass & Diatoms, Cyanobacteria, Green Algae, Chlorophyta & 
    \cellcolor{blue!50}0.937&
    Diatoms, Cyanobacteria, Green Algae &
    \cellcolor{blue!50}0.951\\\hline
\end{longtable}%

\begin{figure}[htbp]
\begin{center}
  \includegraphics[width=0.6\textwidth]{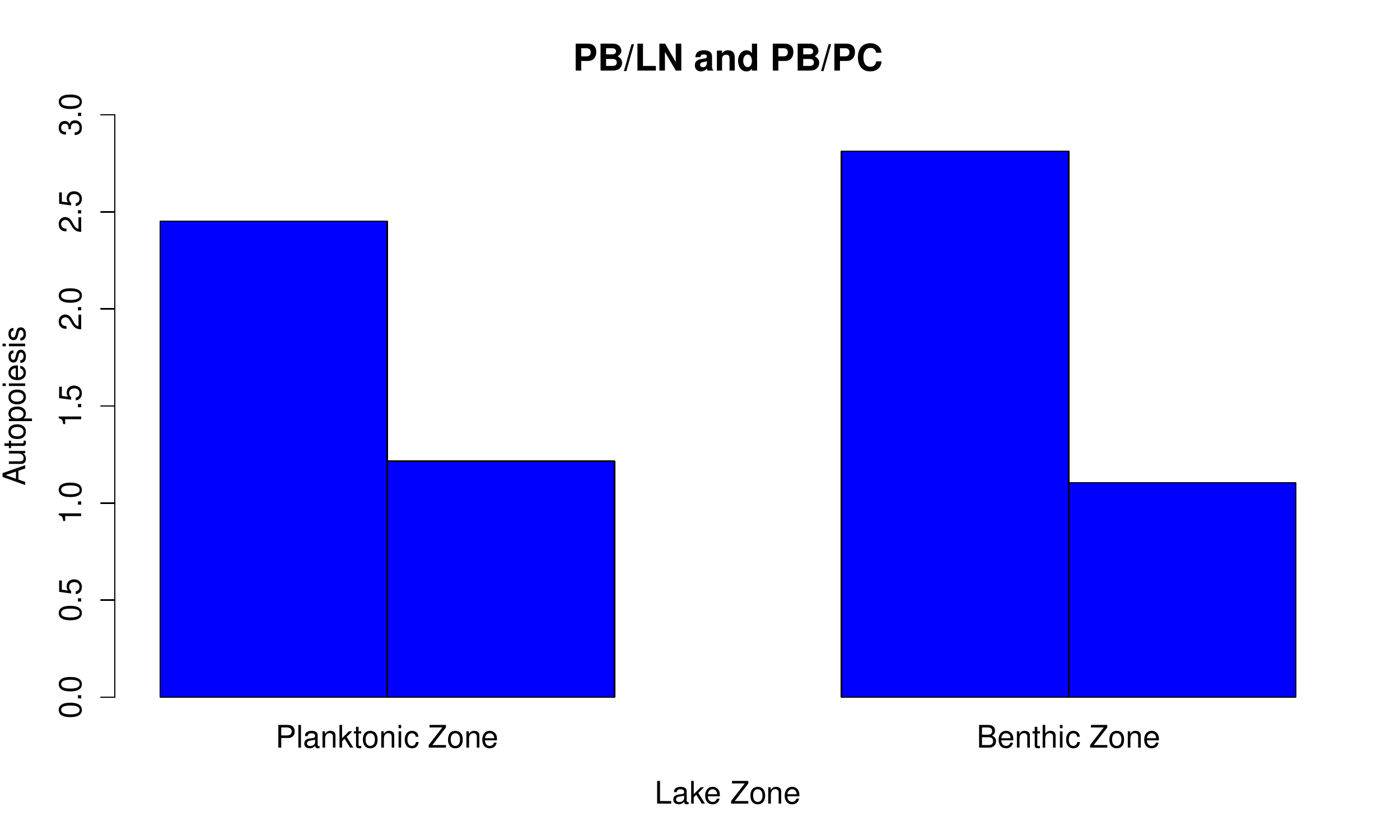}\\
\caption{$A$ of biomass depending on limiting nutrients and physiochemical components.}
\label{fig:A}
\end{center}
\end{figure}

Figure \ref{fig:A} shows the autopoiesis of the two biomass subsystems compared with the $LN$ and $PC$. 
All $A$ values are greater than one. That means that the variables related to living systems have a greater complexity than the variables related to their environment. While we can say that some $PC$ and $LN$ variables have different effects on the planktonic and benthic biomass, we can also estimate that planktonic and benthic biomass are more autonomous compared to their physiochemical and nutrient environments. The very high values of $C$ of biomass imply that these living systems can adapt to the changes of their environments because of the balance between $E$ and $S$ that they have.

\section{Conclusions}
\label{sec:conclusions}

Measuring the complexity of ecological systems has a high potential. Current approaches focus on specific properties of ecosystems. With a general measure, different ecosystems can be compared at different scales, increasing our understanding of ecosystems and complexity itself. 

We applied measures of emergence, self-organization, complexity, homeostasis, and autopoiesis based on information theory to an aquatic ecosystem. The generality and usefulness of the proposed measures will be evaluated gradually, as these are applied to different ecological systems. The potential benefits of general measures as the ones proposed here are manifold. Even if with time more appropriate measures are found, aiming at the goal of finding general measures which can characterize complexity, emergence, self-organization, homeostasis, autopoiesis, and related concepts for any observable ecosystem is a necessary step to make.

\bibliographystyle{splncs03}

\bibliography{carlos,sos,RBN,complex,information,COG,traffic,eco}

\end{document}